\documentstyle[12pt,psfig]{article}
\begin{document}
\begin{center}
\Large{\bf A Second Look at Single Photon Production in $S+Au$ Collisions
 at 200 A$\cdot$GeV and Implications for Quark Hadron Phase Transition}
\vskip 0.2in

\large{Dinesh Kumar Srivastava$^1$  and Bikash Chandra Sinha$^{1,2}$}
\vskip 0.2in

\large{\em $^1$ Variable Energy Cyclotron Centre,\\
 1/AF Bidhan Nagar, Calcutta 700 064, India\\

$^2$ Saha Institute of Nuclear Physics,\\
1/AF Bidhan Nagar, Calcutta 700 064, India\\}

\vskip 0.2in

\large{Erratum: Eur. Phy. J. C 12 (2000) 109.}
Abstract

\vskip 0.2in

\end{center}

It has been recently shown~\cite{MT} that the values of $J_T$ and $J_L$
which appear in Eq.(2) and (5) are too large by a factor of 4,
in the Ref.~\cite{pat}. Correcting for this changes the Fig.~2 (see below).
More-over, now the emissions from the quark matter dominate only for
$p_T <$ 1 GeV/$c$. The basic result of the paper, that the
upper limit of the single-photon production is consistent with the
rates for photon production at two-loop level remains valid.

\setcounter{figure}{1}
\begin{figure}
\psfig{file=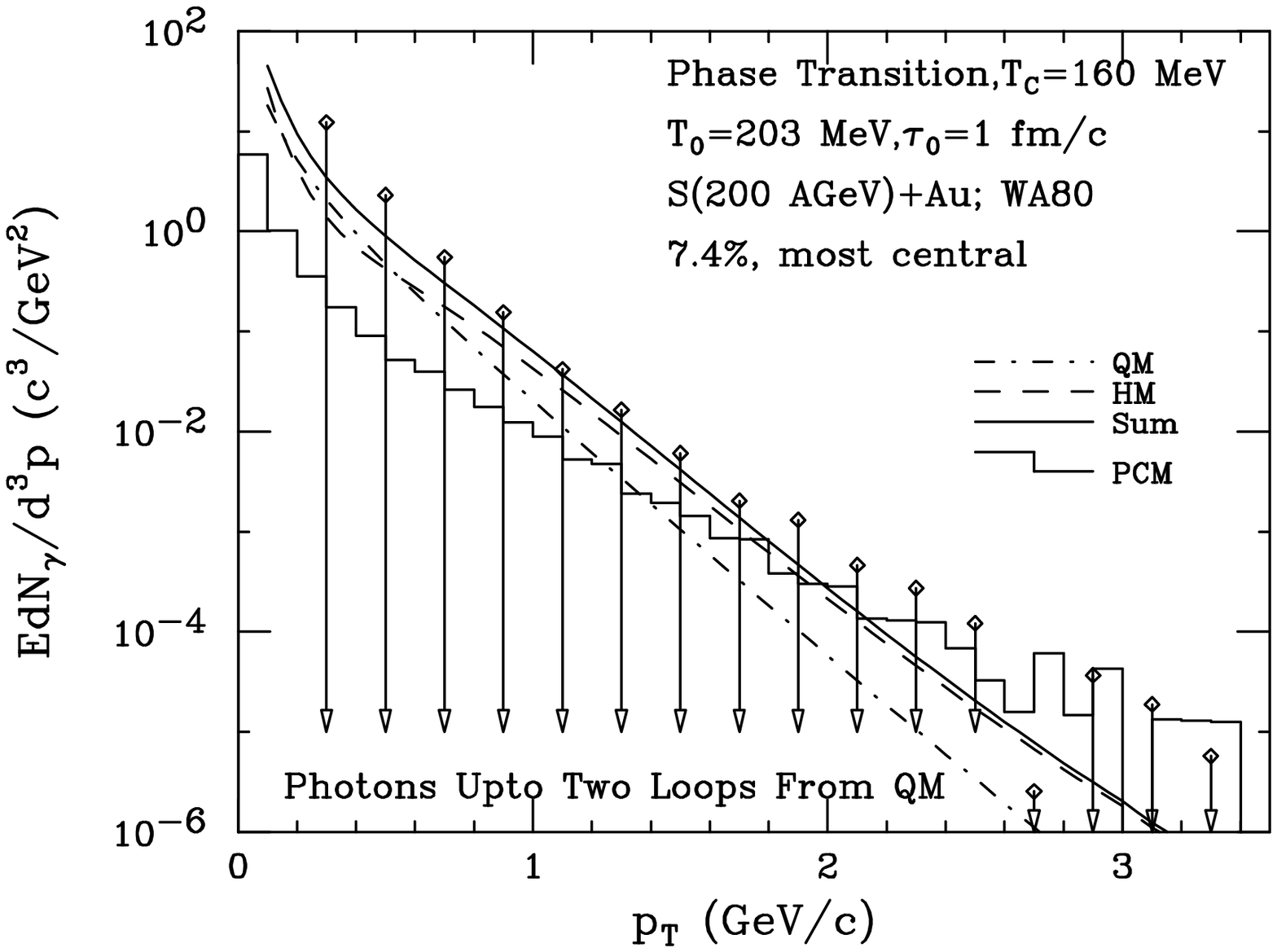,height=12cm,width=15cm}
\vskip 0.1in
\caption{   Same as Fig.~1, with the radiations from the
quark-matter evaluated to the order of two loops. The contribution
of the quark-matter is corrected for the factor of 4 (\protect\cite{MT}).
(This replaces the earlier Fig.~1.)
}
\end{figure}
\end{document}